\documentstyle[axodraw,aps,12pt,epsf]{revtex}

\newcommand\pubnumberk{KA--TP--27--2001}
\newcommand\pubdate{\today}

\newcommand\hepnumber{hep-ph/01xxx}

\def\csumb{
Institut f\"ur Theoretische Physik,
Universit\"at Karlsruhe, \\ D-76128 Karlsruhe, Germany}

\def\Title#1{\begin{center} {\Large\bf #1 } \end{center}}
\def\Author#1{\begin{center}{ \sc #1} \end{center}}
\def\Address#1{\begin{center}{ \it #1} \end{center}}

\newcommand\pubblock{\rightline{\begin{tabular}{l} 
                                                   \pubnumberk\\
         \pubdate\\ \hepnumber \end{tabular}}}
\newenvironment{Abstract}{\begin{quotation}  }{\end{quotation}}

\def\beq{\begin{equation}}
\def\eeq{\end{equation}}
\def\beqn{\begin{eqnarray}}
\def\eeqn{\end{eqnarray}}
\def\bea{\begin{eqnarray}}
\def\eea{\end{eqnarray}}
\def\be{\begin{equation}}
\def\ee{\end{equation}}

\begin{document}
\begin{titlepage}
\pubblock

\Title{ 
Next-to-leading order QCD corrections to
$bg \rightarrow t W^-$ 
at the CERN Large Hadron Collider
}
\vfill
\Author{ Shouhua Zhu 
\footnote{E-mail address: huald@particle.uni-karlsruhe.de}}
\Address{\csumb}
\vfill
\begin{Abstract} 
The next-to-leading order QCD corrections to the Standard
Model process of single top production,
 $bg \rightarrow t W^-$,
 for the CERN Large Hadron Collider
with $\sqrt{s}\!=\!14$ TeV have been calculated. 
For renormalization and factorization scales $\mu=\mu_0$ ($\mu_0=m_t+m_W$), 
the NLO hadronic cross section 
is $\sim 37 $ pb, while
$\sim 25 $ pb for tree level.
The NLO QCD corrections can enhance the cross section by a factor from
$1.33$ to $1.66$ for $\frac{\mu_0}{2}
\ < \ \mu\ <\ 2 \mu_0$. 

\end{Abstract}
\vfill


\end{titlepage}

\eject \baselineskip=0.3in

The top quark plays an important role 
for testing the standard model (SM)
and searching for
 new physics beyond the SM, due to its
large mass, the
same order as the electroweak symmetry breaking scale.
In order to carefully measure the top-quark electroweak interactions
it is useful to consider single top production, in addition to
studying the decay of the top quark in $t \bar t$ events. 
Within the context of the SM, single top production modes provide
a direct measurement of the Cabbibo-Kobayashi-Maskawa matrix
element $V_{tb}$.

At hadron colliders, single top quarks can be produced within the SM
in three different channels, the s-channel $W^*$ production
\cite{Cortese:1991fw,Stelzer:1995mi,Smith:1996ij,Mrenna:1998wp,Heinson:1997zm}, the t-channel
W-exchange mode
\cite{Heinson:1997zm,Dawson:1985gx,Willenbrock:1986cr,Yuan:1990tc,Ellis:1992yw,Bordes:1993sv,Bordes:1995ki,Stelzer:1997ns}, 
and through $t W^-$ production
\cite{Ladinsky:1991ut,Moretti:1997ng}. These three
subprocesses have very different kinematics and experimental
signatures, and are sensitive to different types
of new physics in the top quark sector \cite{Tait:1997fe}.
It should be noticed that the $t W^-$ production
rate is extremely small at Tevatron, but is much greater
at LHC. The study shows \cite{Tait:2000cf} that
a $5 \sigma$ observation of $t W^-$ signal is possible
at very low luminosity at LHC, with $20 fb^{-1}$, cross section
can be measured up to the accuracy of $1\%$.

Both the precise measurement
of $V_{tb}$ and the indirect detection of new physics require an accurate
calculation of the single top quark production cross section. The QCD
corrections to the s-channel $W^*$ production \cite{Smith:1996ij} and 
the t-channel
W-exchange mode \cite{Bordes:1995ki,Stelzer:1997ns} 
have been done. However,  up to now,
only part 
of the QCD corrections [${\cal O} (1/\log m_t^2/m_b^2)$] to the cross
section for $pp \rightarrow $$b g \rightarrow t $$ W^{-}$ 
has been known
\footnote{In Ref. \cite{Giele:1996kr},
the QCD corrections to
 the similar process of $Wc$ production 
at Tevatron has been calculated. } \cite{Tait:2000cf,Giele:1996kr}.
In this letter, the results of the complete 
next-to-leading-order QCD correction to $t W^-$ production will
be presented. A detailed review of the calculation will be 
published elsewhere \cite{FutureBGTW}.

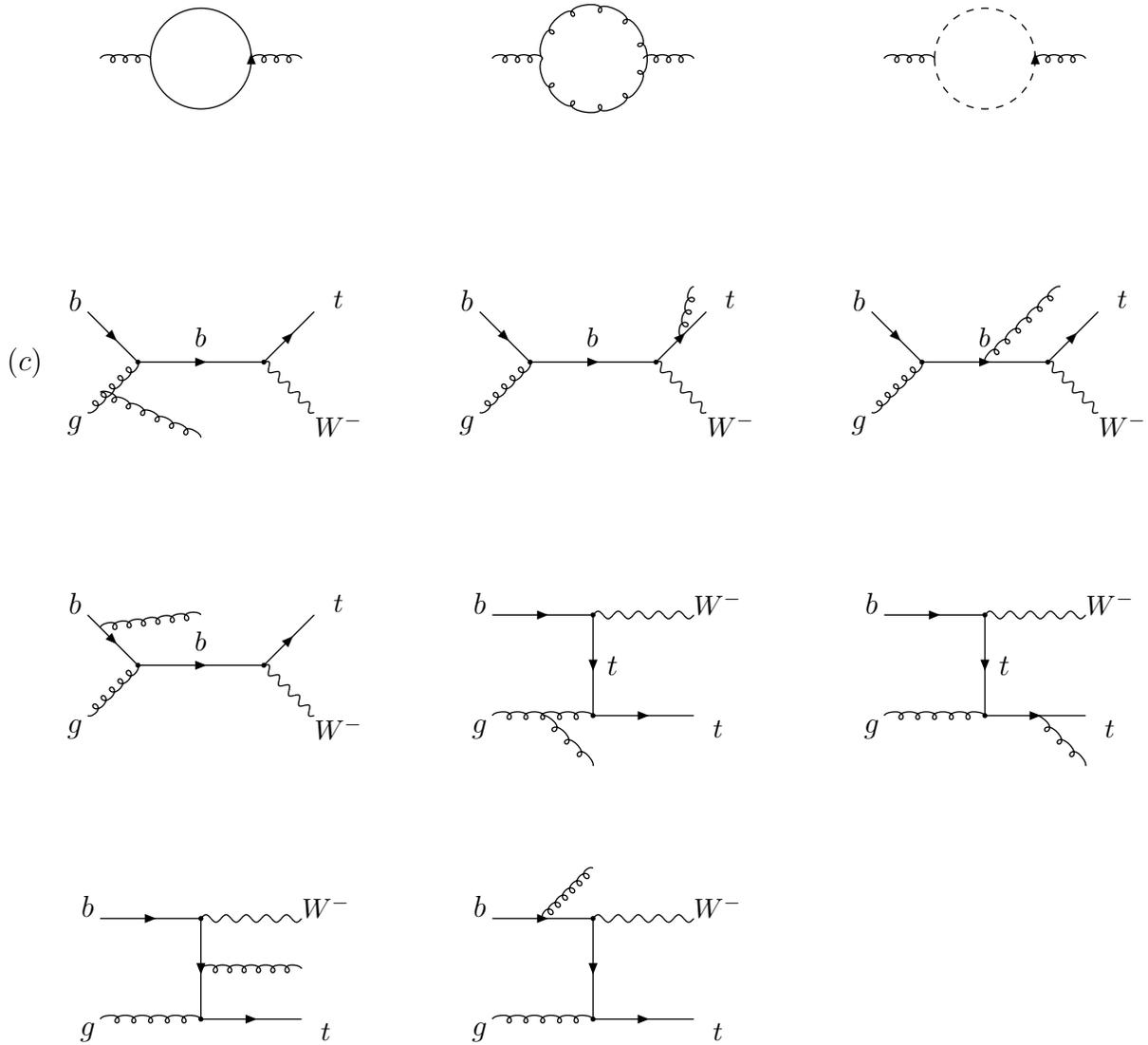
\begin{figure}
\begin{picture}(120,120)(0,0)
\ArrowLine(35,60)(85,60) \ArrowLine(15,80)(35,60)
\Gluon(35,60)(15,40){1.5}{5} \Photon(85,60)(105,40){1.5}{5}
\ArrowLine(85,60)(105,80) \Vertex(35,60){1} \Vertex(85,60){1}
\put(10,85){$b$} \put(10,35){$g$} \put(60,70){\small
$b$} \Text(115,85)[]{\small $t$} \put(115,35){\small
$W^-$} \put(-10,60){$(a)$}
\end{picture}
\hspace{1.0cm}
\begin{picture}(120,120)(0,0)
\ArrowLine(60,80)(60,40) \ArrowLine(20,80)(60,80)
\Gluon(20,40)(60,40){1.5}{5} \Photon(60,80)(100,80){1.5}{5}
\ArrowLine(60,40)(100,40) \Vertex(60,80){1} \Vertex(60,40){1}
\Text(15,85)[]{$b$} \Text(15,35)[]{$g$} \Text(68,60)[]{$t$}
\Text(110,85)[]{\small $W^-$} \Text(110,35)[]{\small $t$}
\end{picture}

\begin{picture}(120,120)(0,0)
\ArrowLine(35,60)(85,60) \ArrowLine(15,80)(35,60)
\Gluon(35,60)(15,40){1.5}{5} \Photon(85,60)(105,40){1.5}{5}
\ArrowLine(85,60)(105,80) \Vertex(35,60){1} \Vertex(85,60){1}
\Text(10,85)[]{$b$} \Text(10,35)[]{$g$} \Text(60,70)[]{\small
$b$} \Text(115,85)[]{\small $t$} \Text(115,35)[]{\small
$W^-$} \Text(-10,60)[]{$(b)$}
\Gluon(20,72)(20,48){1.5}{5}
\end{picture}
\hspace{1.0cm}
\begin{picture}(120,120)(0,0)
\ArrowLine(35,60)(85,60) \ArrowLine(15,80)(35,60)
\Gluon(35,60)(15,40){1.5}{5} \Photon(85,60)(105,40){1.5}{5}
\ArrowLine(85,60)(105,80) \Vertex(35,60){1} \Vertex(85,60){1}
\Text(10,85)[]{$b$} \Text(10,35)[]{$g$} \Text(60,70)[]{\small
$b$} \Text(115,85)[]{\small $t$} \Text(115,35)[]{\small
$W^-$} 
\Gluon(20,72)(60,60){1.5}{5}
\end{picture}
\hspace{1.0cm}
\begin{picture}(120,120)(0,0)
\ArrowLine(35,60)(85,60) \ArrowLine(15,80)(35,60)
\Gluon(35,60)(15,40){1.5}{5} \Photon(85,60)(105,40){1.5}{5}
\ArrowLine(85,60)(105,80) \Vertex(35,60){1} \Vertex(85,60){1}
\Text(10,85)[]{$b$} \Text(10,35)[]{$g$} \Text(60,70)[]{\small
$b$} \Text(115,85)[]{\small $t$} \Text(115,35)[]{\small
$W^-$} 
\Gluon(100,75)(60,60){1.5}{5}
\end{picture}

\begin{picture}(120,120)(0,0)
\ArrowLine(35,60)(85,60) \ArrowLine(15,80)(35,60)
\Gluon(35,60)(15,40){1.5}{5} \Photon(85,60)(105,40){1.5}{5}
\ArrowLine(85,60)(105,80) \Vertex(35,60){1} \Vertex(85,60){1}
\Text(10,85)[]{$b$} \Text(10,35)[]{$g$} \Text(60,70)[]{\small
$b$} \Text(115,85)[]{\small $t$} \Text(115,35)[]{\small
$W^-$} 
\GlueArc(60,60)(15,0,180){1.5}{5}
\end{picture}
\hspace{1.0cm}
\begin{picture}(120,120)(0,0)
\ArrowLine(60,80)(60,40) \ArrowLine(20,80)(60,80)
\Gluon(20,40)(60,40){1.5}{5} \Photon(60,80)(100,80){1.5}{5}
\ArrowLine(60,40)(100,40) \Vertex(60,80){1} \Vertex(60,40){1}
\Text(15,85)[]{$b$} \Text(15,35)[]{$g$} \Text(68,60)[]{$t$}
\Text(110,85)[]{\small $W^-$} \Text(110,35)[]{\small $t$}
\GlueArc(60,60)(15,90,270){1.5}{5}
\end{picture}
\hspace{1.0cm}
\begin{picture}(120,120)(0,0)
\ArrowLine(60,80)(60,40) \ArrowLine(20,80)(60,80)
\Gluon(20,40)(60,40){1.5}{5} \Photon(60,80)(100,80){1.5}{5}
\ArrowLine(60,40)(100,40) \Vertex(60,80){1} \Vertex(60,40){1}
\Text(15,85)[]{$b$} \Text(15,35)[]{$g$} \Text(68,60)[]{$t$}
\Text(110,85)[]{\small $W^-$} \Text(110,35)[]{\small $t$}
\Gluon(40,40)(60,60){1.5}{5}
\end{picture}

\begin{picture}(120,120)(0,0)
\ArrowLine(60,80)(60,40) \ArrowLine(20,80)(60,80)
\Gluon(20,40)(60,40){1.5}{5} \Photon(60,80)(100,80){1.5}{5}
\ArrowLine(60,40)(100,40) \Vertex(60,80){1} \Vertex(60,40){1}
\Text(15,85)[]{$b$} \Text(15,35)[]{$g$} \Text(68,60)[]{$t$}
\Text(110,85)[]{\small $W^-$} \Text(110,35)[]{\small $t$}
\Gluon(80,40)(60,60){1.5}{5}
\end{picture}
\hspace{1.0cm}
\begin{picture}(120,120)(0,0)
\ArrowLine(60,80)(60,40) \ArrowLine(20,80)(60,80)
\Gluon(20,40)(60,40){1.5}{5} \Photon(60,80)(100,80){1.5}{5}
\ArrowLine(60,40)(100,40) \Vertex(60,80){1} \Vertex(60,40){1}
\Text(15,85)[]{$b$} \Text(15,35)[]{$g$} \Text(68,60)[]{$t$}
\Text(110,85)[]{\small $W^-$} \Text(110,35)[]{\small $t$}
\Gluon(40,80)(60,60){1.5}{5}
\end{picture}
\hspace{1.0cm}
\begin{picture}(120,120)(0,0)
\ArrowLine(60,80)(60,40) \ArrowLine(20,80)(60,80)
\Gluon(20,40)(60,40){1.5}{5} \Photon(60,80)(100,80){1.5}{5}
\ArrowLine(60,40)(100,40) \Vertex(60,80){1} \Vertex(60,40){1}
\Text(15,85)[]{$b$} \Text(15,35)[]{$g$} \Text(68,60)[]{$t$}
\Text(110,85)[]{\small $W^-$} \Text(110,35)[]{\small $t$}
\Gluon(40,80)(40,40){1.5}{5}
\end{picture}

\begin{picture}(120,120)(0,0)
\ArrowLine(60,80)(60,40) \ArrowLine(20,80)(60,80)
\Gluon(20,40)(60,40){1.5}{5} \Photon(60,80)(100,80){1.5}{5}
\ArrowLine(60,40)(100,40) \Vertex(60,80){1} \Vertex(60,40){1}
\Text(15,85)[]{$b$} \Text(15,35)[]{$g$} \Text(68,60)[]{$t$}
\Text(110,85)[]{\small $W^-$} \Text(110,35)[]{\small $t$}
\Gluon(40,80)(80,40){1.5}{5}
\end{picture}
\hspace{1.0cm}
\begin{picture}(120,120)(0,0)
\ArrowLine(60,40)(60,80) \ArrowLine(20,80)(40,80)
\Gluon(20,40)(40,40){1.5}{3} \Photon(60,40)(100,40){1.5}{5}
\ArrowLine(60,80)(100,80) \Vertex(60,80){1} \Vertex(60,40){1}
\Text(15,85)[]{$b$} \Text(15,35)[]{$g$} \Text(68,60)[]{$t$}
\Text(110,85)[]{\small $t$} \Text(110,35)[]{\small $W^-$}
\ArrowLine(40,80)(40,40)
\ArrowLine(40,40)(60,40)
\Gluon(40,80)(60,80){1.5}{3}
\end{picture}
\hspace{1.0cm}
\begin{picture}(120,120)(0,0)
\Line(40,60)(80,60) \ArrowLine(20,60)(40,60)
\ArrowLine(80,60)(100,60) \GlueArc(60,60)(20,0,180){1.5}{5}
\Vertex(40,60){1} \Vertex(80,60){1} \Text(20,50)[]{$t(b)$}
\Text(100,50)[]{$t(b)$} 
\end{picture}

\begin{picture}(120,120)(0,0)
\Gluon(20,60)(40,60){1.5}{3}
\Gluon(80,60)(100,60){1.5}{3}
\ArrowArc(60,60)(20,-180,180)
\end{picture}
\hspace{1.0cm}
\begin{picture}(120,120)(0,0)
\Gluon(20,60)(40,60){1.5}{3}
\Gluon(80,60)(100,60){1.5}{3}
\GlueArc(60,60)(20,-180,180){1.5}{10}
\end{picture}
\hspace{1.0cm}
\begin{picture}(120,120)(0,0)
\Gluon(20,60)(40,60){1.5}{3}
\Gluon(80,60)(100,60){1.5}{3}
\DashArrowArc(60,60)(20,-180,180){3}
\end{picture}


\begin{picture}(120,120)(0,0)
\ArrowLine(35,60)(85,60) \ArrowLine(15,80)(35,60)
\Gluon(35,60)(15,40){1.5}{5} \Photon(85,60)(105,40){1.5}{5}
\ArrowLine(85,60)(105,80) \Vertex(35,60){1} \Vertex(85,60){1}
\Text(10,85)[]{$b$} \Text(10,35)[]{$g$} \Text(60,70)[]{\small
$b$} \Text(115,85)[]{\small $t$} \Text(115,35)[]{\small
$W^-$} \Text(-10,60)[]{$(c)$}
\Gluon(20,48)(60,30){1.5}{6}
\end{picture}
\hspace{1.0cm}
\begin{picture}(120,120)(0,0)
\ArrowLine(35,60)(85,60) \ArrowLine(15,80)(35,60)
\Gluon(35,60)(15,40){1.5}{5} \Photon(85,60)(105,40){1.5}{5}
\ArrowLine(85,60)(105,80) \Vertex(35,60){1} \Vertex(85,60){1}
\Text(10,85)[]{$b$} \Text(10,35)[]{$g$} \Text(60,70)[]{\small
$b$} \Text(115,85)[]{\small $t$} \Text(115,35)[]{\small
$W^-$} 
\Gluon(95,70)(100,90){1.5}{3}
\end{picture}
\hspace{1.0cm}
\begin{picture}(120,120)(0,0)
\ArrowLine(35,60)(85,60) \ArrowLine(15,80)(35,60)
\Gluon(35,60)(15,40){1.5}{5} \Photon(85,60)(105,40){1.5}{5}
\ArrowLine(85,60)(105,80) \Vertex(35,60){1} \Vertex(85,60){1}
\Text(10,85)[]{$b$} \Text(10,35)[]{$g$} \Text(60,70)[]{\small
$b$} \Text(115,85)[]{\small $t$} \Text(115,35)[]{\small
$W^-$} 
\Gluon(60,60)(90,90){1.5}{6}
\end{picture}

\begin{picture}(120,120)(0,0)
\ArrowLine(35,60)(85,60) \ArrowLine(15,80)(35,60)
\Gluon(35,60)(15,40){1.5}{5} \Photon(85,60)(105,40){1.5}{5}
\ArrowLine(85,60)(105,80) \Vertex(35,60){1} \Vertex(85,60){1}
\Text(10,85)[]{$b$} \Text(10,35)[]{$g$} \Text(60,70)[]{\small
$b$} \Text(115,85)[]{\small $t$} \Text(115,35)[]{\small
$W^-$} 
\Gluon(20,75)(60,80){1.5}{6}
\end{picture}
\hspace{1.0cm}
\begin{picture}(120,120)(0,0)
\ArrowLine(60,80)(60,40) \ArrowLine(20,80)(60,80)
\Gluon(20,40)(60,40){1.5}{5} \Photon(60,80)(100,80){1.5}{5}
\ArrowLine(60,40)(100,40) \Vertex(60,80){1} \Vertex(60,40){1}
\Text(15,85)[]{$b$} \Text(15,35)[]{$g$} \Text(68,60)[]{$t$}
\Text(110,85)[]{\small $W^-$} \Text(110,35)[]{\small $t$}
\Gluon(40,40)(60,20){1.5}{3}
\end{picture}
\hspace{1.0cm}
\begin{picture}(120,120)(0,0)
\ArrowLine(60,80)(60,40) \ArrowLine(20,80)(60,80)
\Gluon(20,40)(60,40){1.5}{5} \Photon(60,80)(100,80){1.5}{5}
\ArrowLine(60,40)(100,40) \Vertex(60,80){1} \Vertex(60,40){1}
\Text(15,85)[]{$b$} \Text(15,35)[]{$g$} \Text(68,60)[]{$t$}
\Text(110,85)[]{\small $W^-$} \Text(110,35)[]{\small $t$}
\Gluon(80,40)(100,20){1.5}{3}
\end{picture}

\begin{picture}(120,120)(0,0)
\ArrowLine(60,80)(60,40) \ArrowLine(20,80)(60,80)
\Gluon(20,40)(60,40){1.5}{5} \Photon(60,80)(100,80){1.5}{5}
\ArrowLine(60,40)(100,40) \Vertex(60,80){1} \Vertex(60,40){1}
\Text(15,85)[]{$b$} \Text(15,35)[]{$g$} 
\Text(110,85)[]{\small $W^-$} 
\Text(110,35)[]{\small $t$}
\Gluon(60,60)(100,60){1.5}{6}
\end{picture}
\hspace{1.0cm}
\begin{picture}(120,120)(0,0)
\ArrowLine(60,80)(60,40) \ArrowLine(20,80)(60,80)
\Gluon(20,40)(60,40){1.5}{5} \Photon(60,80)(100,80){1.5}{5}
\ArrowLine(60,40)(100,40) \Vertex(60,80){1} \Vertex(60,40){1}
\Text(15,85)[]{$b$} \Text(15,35)[]{$g$} 
\Text(110,85)[]{\small $W^-$} 
\Text(110,35)[]{\small $t$}
\Gluon(40,80)(60,100){1.5}{6}
\end{picture}

\caption{Feynmann diagrams for $bg \rightarrow t W^-$: the Born level (a),
the virtual gluon exchange (b) and the gluon radiation (c).  }
\label{fig1}
\end{figure}

The Feynman diagrams for $tW^-$ production via
the parton process
$b(p_1) g(p_2)\rightarrow t (k_1) W^{-}(k_2)$,
including the QCD corrections, are shown in Fig.~1.
The Born diagrams are shown in
Fig. \ref{fig1}(a), the NLO diagrams by virtual
gluon-exchange and
the gluon-radiation (gr)
are shown in  Fig. \ref{fig1}(b) 
and Fig. \ref{fig1}(c).
Because the topologies of 
the initial-gluon (ig) diagrams are the same with 
the gluon-radiation processes, we don't show here the
diagrams, which can be easily obtained from
Fig. \ref{fig1}(c) by treating two gluons as initial partons. 
The diagrams are created by use of
FeynArts~\cite{Kublbeck:1990xc,Hahn:1999wr} and are handled with the help of
FeynCalc~\cite{Mertig:1991an}. We perform all the calculations in
$d=4-2\epsilon$ dimensions and adopt
$\overline{MS}$ renormalization and factorization schemes.

The total cross section for $pp \rightarrow b g \rightarrow t
  W^-$ at ${\cal O}(\alpha_s^2)$ can be written as:
\begin{eqnarray}
\sigma (s)  &=& \sigma^0 (s) +\sigma^{vir}(s) +\sigma^{gr}(s)+\sigma^{ig}(s), 
\nonumber \\
&\equiv& \sigma^{2\ body} (s)+\sigma^{gr}(s)+\sigma^{ig}(s), 
\end{eqnarray}
where
\begin{eqnarray}
\sigma^{2\ body} (s) &=& \int^1_{z_0} dz \left( \frac{d L}{d z} \right)_{bg}
\hat \sigma^{2\ body} (z^2 s, \mu_r), \nonumber \\
\left( \frac{d L}{d z} \right)_{bg}
&=& 2 z \int^1_{z^2} \frac{dx}{x} f_{b/P} (x,\mu_f) 
f_{g/P} (\frac{z^2}{x},\mu_f)
\label{eqnlumi}
\end{eqnarray}
with $z_0=(m_W+m_t)/\sqrt{s}$.
Here $\sigma^0$, $\sigma^{vir}$, $\sigma^{gr}$ and
$\sigma^{ig}$ are contributions from tree level, virtual,
gluon-radiation and initial-gluon diagrams. 
The two-body subprocess cross section can be expressed as
\begin{eqnarray}
\hat \sigma^{2\ body} (\hat s)
&=& \int 
\overline{\sum} |M_{ren}|^2 d \Phi_2 \nonumber \\
 &=&\int  \overline{\sum} |M_{0}|^2 d \Phi_2  +
\int \overline{\sum} 2 {\rm Re} (M^{vir} M_0^+) d \Phi_2 \nonumber \\
&\equiv& \hat \sigma^0 + \hat \sigma^{vir}.
\end{eqnarray}
Here $M_0$, $M_{ren}$ and $ d \Phi_2$ are the tree level amplitude,
the renormalized amplitude and
the two-body phase space in $d$ dimension. 
The details of the renormalization procedure and the explicit
expressions of $M_{ren}$ will be given in Ref. \cite{FutureBGTW}.
As usual, $\hat \sigma^{vir}$ contains infrared
divergences after renormalization, 
which can only be cancelled by adding
contributions from $\sigma^{gr}$. The remaining collinear divergences
are absorbed by the redefinition of the parton distribution functions (PDF).

The real corrections $\sigma^{gr}$ and
$\sigma^{ig}$ have been computed using the two
cut-off phase space slicing method (TCPSSM)
\cite{Harris:2001sx}. The main idea of TCPSSM is
to introduce two small constants $\delta_s$, $\delta_c$.
The three-body
phase space can then be divided into soft and hard regions
according to parameter $\delta_s$, and the hard region
is further divided into collinear and non-collinear
regions according to parameter $\delta_c$. 
In the soft and collinear regions, approximations
can be made and analytical results can be easily
obtained. In the non-collinear region, numerical results
can be calculated in four dimension by standard Monte Carlo packages because
it contains no divergences. The physical results
should be independent on these artificial
parameters $\delta_s$ and $\delta_c$, which offers a crucial 
way to check our results. 

We can write the $\sigma^{gr}$ as
\begin{eqnarray}
\sigma^{gr}= \sigma^{gr}_s+ \sigma^{gr}_{c} +
\sigma^{gr}_{fin},
\end{eqnarray}
where $\sigma^{gr}_s$, $\sigma^{gr}_{c}$ and $\sigma^{gr}_{fin}$
are the contributions in the soft, collinear and non-collinear regions. 
In the soft region, we can write the $\sigma^{gr}_s$ as \cite{Harris:2001sx}
\begin{eqnarray}
\sigma^{gr}_s &=& \int^1_{z_0} dz \left( \frac{d L}{d z} \right)_{bg}
\hat \sigma_{s}^{gr} (z^2 s, \mu_r), \nonumber \\
\hat \sigma_{s}^{gr} &=& \hat{\sigma}^0 \left[ \frac{\alpha_s}{2\pi}
\frac{\Gamma(1-\epsilon)}{\Gamma(1-2\epsilon)} \left( \frac{4\pi\mu_r^2}{\hat 
s} \right)^\epsilon \right]
\left( \frac{A_2^s}{\epsilon^2}+\frac{A_1^s}{\epsilon}+A_0^s \right).
\end{eqnarray}
The lengthy expressions of the coefficients $A_i$ will be given
in Ref. \cite{FutureBGTW}. 
In the collinear region,  the
contributions after factorization can be written as 
two parts \cite{Harris:2001sx}
\begin{eqnarray}
\sigma^{gr}_{c}&=& \int^1_{z_0} dz \hat \sigma^0 \left\{
\left( \frac{d L}{d z} \right)_{bg}
\hat \sigma_c^{gr}+ \left[ 
\left( \frac{d L}{d z} \right)^{gr}_{b\tilde{g}}+
\left( \frac{d L}{d z} \right)^{gr}_{g\tilde{b}} \right]  \right\},
\end{eqnarray}
where the definition of the luminosity is similar to that in
 Eq. \ref{eqnlumi}. 
Here the $\tilde{g}$ and $\tilde{b}$ are
\begin{eqnarray}
\tilde{g}/\tilde{b}(x,\mu_f) =  \int_x^{1-\delta_s} \frac{dy}{y}
                       f_g/f_b(x/y,\mu_f) \tilde{P}_{gg/bb}(y)
\end{eqnarray}
with
\begin{eqnarray}
\tilde{P}_{ij}(y) = P_{ij}(y)\log\left(\delta_c\frac{1-y}{y}
\frac{\hat s}{\mu_f^2}\right) - P_{ij}^{\prime}(y) \,,
\label{eq8}
\end{eqnarray}
where $\hat s$ is the subprocess center-of-mass energy and
\begin{eqnarray}
P_{bb}(z) &=& C_F \frac{1+z^2}{1-z}  \nonumber \\
P_{bb}^{\prime}(z) &=& -C_F(1-z)  \nonumber \\
P_{gg}(z) &=& 2N\left[ \frac{z}{1-z}+\frac{1-z}{z}+z(1-z)\right] 
\nonumber \\
P_{gg}^{\prime}(z) &=& 0 
\end{eqnarray}
with $N=3$ and $C_F=4/3$. $\hat \sigma_c^{gr}$ can be written as
\begin{eqnarray}
\hat \sigma_c^{gr} &=& 
\left[ \frac{\alpha_s}{2\pi} \frac{\Gamma(1-\epsilon)}{\Gamma(1-2\epsilon)}
\left(\frac{4 \pi \mu_r^2}{\hat s}\right)^{\epsilon}\right]
 \left\{
 \frac{A_1^{sc}(b\rightarrow bg)}{\epsilon}+
 \frac{A_1^{sc}(g\rightarrow gg)}{\epsilon}
+ A_0^{sc}(b\rightarrow bg)+
A_0^{sc}(g\rightarrow gg)
\right\},
\end{eqnarray}
where
\begin{eqnarray}
A_0^{sc} &=& A_1^{sc} \log \left( \frac{\hat s}{\mu_f^2} \right)
 \\
A_1^{sc}(b\rightarrow bg) &=& C_F(2 \log \delta_s + 3/2 )  \\
A_1^{sc}(g\rightarrow gg) &=& 2N \log \delta_s + (11N-2 n_f)/6 \, 
\end{eqnarray}
with $n_f=5$.

For the initial-gluon processes, the results are much simpler
compared to the gluon radiation processes, 
\begin{eqnarray}
\sigma^{ig}= \sigma^{ig}_{c} + \sigma^{ig}_{fin},
\end{eqnarray}
where $\sigma^{ig}_{c}$ and $\sigma^{ig}_{fin}$
are the contributions in the collinear and the non-collinear regions. 
 After factorization,
 we can write $\sigma^{ig}_{c}$ as \cite{Harris:2001sx}
\begin{eqnarray}
\sigma^{ig}_{c}&=& \int^1_{z_0} dz \hat \sigma^0 
\left( \frac{d L}{d z} \right)^{ig}_{g\tilde b},
\end{eqnarray}
where the definition of $\tilde b$ in 
the luminosity $\left( \frac{d L}{d z} \right)^{ig}_{g\tilde b} $ is 
\begin{eqnarray}
\tilde{b}(x,\mu_f) =  \int_x^1 \frac{dy}{y}
                       f_g(x/y,\mu_f) \tilde{P}_{bg}(y).
\end{eqnarray}
The splitting  functions in $\tilde{P}_{bg}$, defined in
Eq. \ref{eq8}, contains the parts 
\begin{eqnarray}
P_{bg}(z) &=& \frac{1}{2} \left[ z^2+(1-z)^2 \right]  \nonumber \\
P_{bg}^{\prime}(z) &=& -z(1-z). 
\end{eqnarray}

Another important issue, which does not exist in
gluon-radiation processes,  is the procedure how to
subtract the contribution of the on-shell anti-top quark decay to $W^-$ and 
$\bar b$
from $\sigma^{ig}$, besides subtracting double counting of
$g\rightarrow b\bar b$ in collinear region.  As in Ref. 
\cite{Tait:2000cf}, in order to remove all of the
$\bar t$ contribution, we should subtract the term given by
(in the narrow decay width limit)  
\begin{eqnarray}
\sigma&=&\sigma^{LO}(gg\rightarrow t \bar t) {\cal B} (\bar t \rightarrow
W^- \bar b), 
\end{eqnarray}
where $\sigma^{LO}(gg\rightarrow t \bar t)$ and
${\cal B} (\bar t \rightarrow
W^- \bar b)$ 
are LO cross section of $gg\rightarrow t \bar t$ and
branching ratio of the decay $\bar t \rightarrow
W^- \bar b$. 

It should be noticed that at $O(\alpha_s^2)$, there are another
QCD corrections arising from
$q \bar q \rightarrow t \bar b W^-$ and $b q (\bar q)
\rightarrow t W^- q (\bar q)$ $[q=u, d, s]$, which
can be treated by the similar methods described above.
However, due to the lower luminosity of $q \bar q$ and
$b q (\bar q)$ compared to those of $gg$ and $bg$ at the LHC, the
QCD corrections are smaller, which are only a few percents.
For completeness, we will include the initiated light quark
contributions in our numerical results.

Our numerical results are obtained using CTEQ5M (CTEQ5L) PDF
\cite{Lai:2000wy} for NLO (LO) cross-section calculations.  
The $2$-loop ($1$-loop) evolution of $\alpha_s(\mu)$ is adopted 
for NLO (LO) calculation and 
$\Lambda^{(5)}=226$ $(146)$ MeV for two-loop (one-loop) evolution. The top-quark
pole mass is taken to be $m_t\!=\!175$~GeV; 
for simplicity, the bottom-quark mass
has been omitted, and the renormalization and factorization scales
are taken to be the same. We have compared the numerical results
of the initial-gluon contribution to that in Ref. \cite{Tait:2000cf},
and both results are in good agreement.

\begin{figure}
\begin{center}
\epsfxsize=15cm \epsfbox{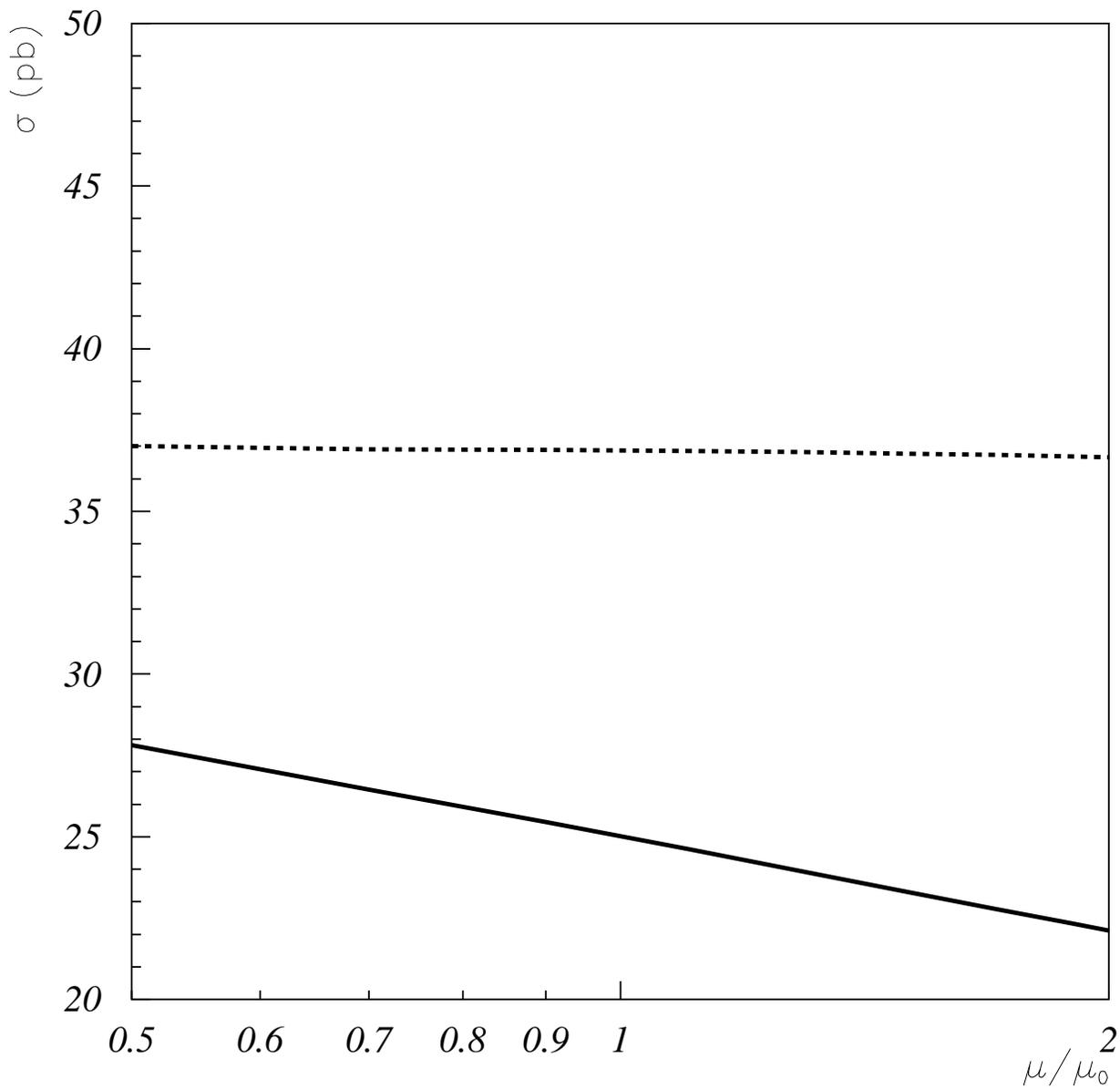}
\caption[ ]{ 
Cross sections of NLO (dashed) and LO (solid) for
  $pp \rightarrow b  g \rightarrow t W^-$ as functions of
  $\mu/\mu_0$ at the LHC with $\sqrt{s}=14$
TeV, where $\mu_0=m_t+m_W$.  }
\label{fig2}
\end{center}
\end{figure}

\begin{figure}
\begin{center}
\epsfxsize=15cm \epsfbox{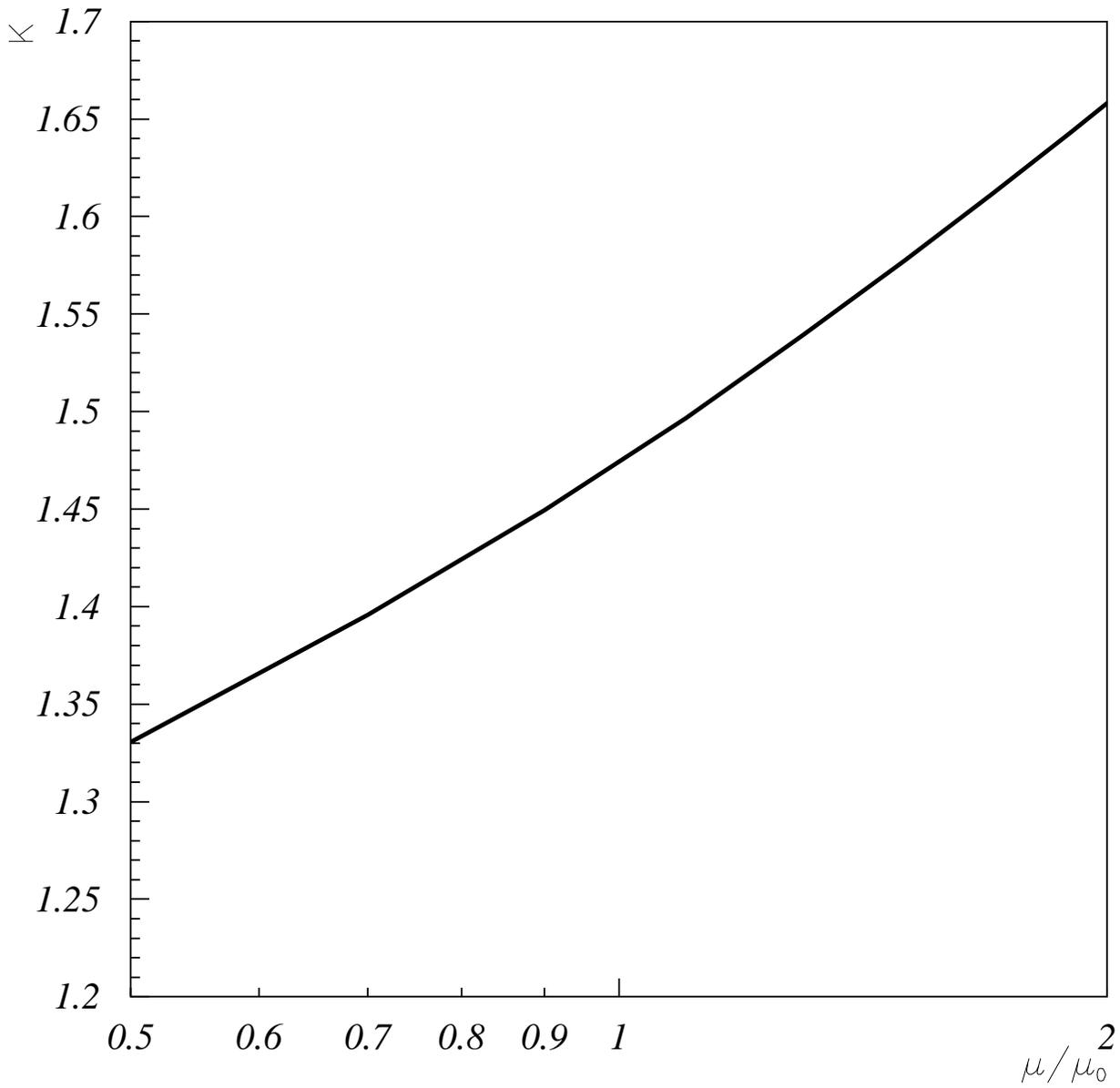}
\caption[ ]{ K factor
for $pp \rightarrow bg \rightarrow t W^-$ 
  as functions of
  $\mu/\mu_0$ at the LHC.  }
\label{fig3}
\end{center}
\end{figure}

In Fig.~\ref{fig2} we show the tree level and NLO cross sections
as a function of renormalization and factorization scales $\mu/\mu_0$
($\mu_0=m_t+m_W$).
From the figure we can see that 
the NLO result is greater than the lowest order one. The NLO cross section
is $\sim 37 $ pb when $\mu=\mu_0$, while 
$\sim 25 $ pb at tree level.  
In Fig.~\ref{fig3},
the K factor (defined as the ratio of the NLO
cross section to the LO one)
is shown.  For 
$\frac{\mu_0}{2} \ < \mu
 \  < 2 \mu_0$, 
the K factor varies
roughly between $1.33$ and $1.66$.  


To summarize, the next-to-leading order QCD corrections to the Standard
Model process $bg \rightarrow t W^-$ at CERN large hadron collider with
$\sqrt{s} = 14$ TeV have been calculated. 
The NLO QCD corrections can enhance the cross section by a factor from
$1.33$ to $ 1.66$ for renormalization and factorization scales $\frac{\mu_0}{2}
\ < \ \mu\ <\ 2 \mu_0$. We should note here that the results presented
in this letter
are for the process $bg \rightarrow t W^-$; they are 
the same for the charge conjugate process $\bar b g \rightarrow W^+\bar t$.
\\

The author would like to thank Prof. W. Hollik and Prof. C.S. Li
for stimulating discussions.
This work was supported in part by the Alexander von Humboldt
Foundation and
National Nature Science Foundation of China.
Parts of the calculations have been performed on the QCM
cluster at the University of Karlsruhe, supported by the
DFG-Forschergruppe ''Quantenfeldtheorie, Computeralgebra und
Monte-Carlo-Simulation''.

\newpage

\end{document}